\documentclass[12pt]{article}
\usepackage{epsfig}
\begin{document}

\begin{small}
\begin{center}

{\large\bf Study of the $\psi(2S)$ decay to
$p\overline{p}\pi^+\pi^-(\pi^0)$}
\vspace{0.6cm}

J.~G.~Bian
\\
\vspace{0.2cm}
\vspace{0.2cm}
{\it
Institute of High Energy Physics, Beijing 100049, People's 
Republic of China}\\

\end{center}
\end{small}

\parbox[c]{17.0truecm}{
{\hskip 0.6cm}The branching ratios of $\psi(2S)\rightarrow p\overline{p}
\pi^+\pi^-$ and $\psi(2S)\rightarrow p\overline{p}\omega,~\omega\rightarrow
\pi^+\pi^-\pi^0$ are measured
and the $\pi^+\pi^-$ invariant mass distribution
in the first decay is discussed by analysing 
14 million produced $\psi(2S)$ events collected by
the BESII detector at the BEPC. 
}
\\
\\
\parbox[c]{17.0truecm}{
\mbox{}{\hskip 4cm}PACS numbers: 13.20.Gv, 13.65.+i, 
13.75.Lb, 14.40.Cs. }
\\
\\
\mbox{}{\hskip 0.6cm}

In the quarkonium model, the $J/\psi$ and
$\psi(2S)$ are respectively the ground state$^{[1]}$ and
 the first radial excitation of the $^3S $
$c\overline{c}$ bound state$^{[2]}$. As such,
their decay is supposed to have a similar feature.
In both $J/\psi$ decay and $\psi(2S)$ decay, 
the $p\overline{p}\omega$ and
$p\overline{p}\rho^0$ are allowed three body states. 
For $J/\psi$ decay,
the production of $p\overline{p}\omega$ is observed
at the fair fraction$^{[3]}$ of $(1.30\pm0.25)\times 10^{-3}$
 relative to other allowed three body
states, while only a upper limit is set to
the branching ratio$^{[4]}$ of the production of $p\overline{p}\rho^0$.
 Therefore, to observe  how the $p\overline{p}\omega$
$p\overline{p}\rho^0$ 
are produced in $\psi(2S)$ decay
is very interesting. 

The $\omega$ decays to $\pi^+\pi^-$ and $\pi^+\pi^-\pi^0$
at $1.7\%$  and $89.7\%$ branching ratios$^{[4,5]}$.
The $p\overline{p}\omega$ is observed by BESI$^{[6]}$. 
In the paper, we present the $\pi^+\pi^-$ mass spectrum
in $\psi(2S)$ decay to $p\overline{p}\pi^+\pi^-$
and the $\pi^+\pi^-\pi^0$ mass spectrum
 in $\psi(2S)$ decay to $p\overline{p}\pi^+\pi^-\pi^0$.
The analysis is based on
14 million produced $\psi(2S)$ events collected by
the BESII detector at the BEPC.

This study shows an unknown bump of about 20 MeV width at 727 MeV
 in $\pi^+\pi^-$ mass spectra. 
The BES apparatus is a magnetic
spectrometer working at $e^+e^-$ colliding mode, which has been fully 
described elsewhere$^{[7]}$.

 
The decay of $\psi(2S)$ to $p\overline{p}\pi^+\pi^-\pi^0$
is discussed
 fisrt beacuse of $\omega$ decays to
$\pi^+\pi^-\pi^0$ in large branching ratio. 

 Candidates for
 $\psi(2S)\rightarrow p\overline{p}\pi^+\pi^-\pi^0 $ events are
selected by requiring exactly four reconstructed
charged tracks in the drift
chamber with zero net charge. Tracks with 
transverse momentum $p_{xy}>0.08$ GeV and
$\mid cos\theta_{ch} \mid<0.85$  are accepted,
where $\theta_{ch}$ is the polar angle
with respect to the beam direction. The particles are
identified by  requiring that their combination weights of
 time-of-flight (TOF) and the ionization energy loss (dE/d{\it x})
in the drift chamber
be consistent with the corresponding particle  hypothesis.
The events with at least two charged particle satisfying
proton and anti-proton   hypotheses
and one satisfying pion  hypothesis are selected.

To remove the contamination from
$\psi(2S)\rightarrow \Lambda\overline{\Lambda}\pi^0$,
$M_{p\pi^-}>1.15$ $GeV$ and
$M_{\overline{p}\pi^+}>1.15$ $GeV$ are required.

An isolated  photon is defined as a
cluster in the barrel shower counter  with
at least two readout layers hit,
energy larger than 30 MeV,
outside  a $25^\circ$ cone
 around the $\overline{p}$ and outside a $12^{\circ}$ cone around
other charged particles  to reject
possible fake photons produced by $\overline{p}$ annihilation
and/or radiated by other charged particles inside the shower
counter.
The cluster is also required to has its
incident direction (from the interaction point to
the first hit point in BSC)
inside a $20^{\circ}$ cone
around the cluster developing direction.

The events are kinematically fitted
for the $\psi(2S)$ $\rightarrow p\overline{p}\pi^+\pi^-\pi^0$
topology by imposing
energy and momentum constraints (4C).
$\chi^2_{p\overline{p}\pi^+\pi^-\gamma\gamma}<20$ and 
$\chi^2_{p\overline{p}\pi^+\pi^-\gamma\gamma}<
\chi^2_{p\overline{p}\pi^+\pi^-\gamma}$
to remove $\psi(2S)\rightarrow p\overline{p}\pi^+\pi^-\gamma$.

The combination with the smallest $\chi^2$ in the 4C fit is chosen to
identify the radiative photon if there are more than two
photon candidates in an event.
Fig. 1 showes the $\gamma\gamma$ invariant mass distribution.
The $\pi^0$  signal is seen at 0.135 GeV. 
To select $\pi^0$, it is required
$\mid M_{\gamma\gamma}-0.135\mid<0.040$GeV $(3\sigma)$.

A contamination is $\psi(2S)\rightarrow \pi^+\pi^-J/\psi,
J/\psi\rightarrow p\overline{p}\pi^0$. To remove it
it is required
$\mid M_{p \overline{p}\pi^0}-3.097\mid>0.065 $ $GeV$ $(3\sigma)$.
(why  do not use $M_{p \overline{p}\pi^0}<3.0$ $GeV$,
because $M_{p \overline{p}\pi^0}$ distribution is around
3.0 $GeV$ for $\psi(2S)\rightarrow 
p\overline{p}\omega,\omega\rightarrow \pi^+\pi^-\pi^0$.)

\begin{center}
 \psfull
\epsfig{file=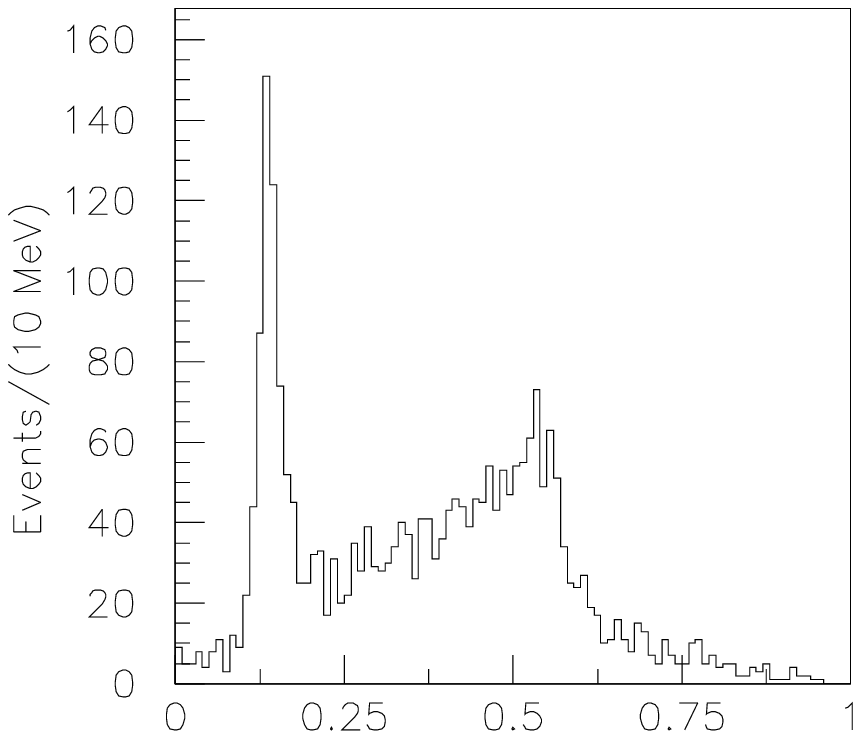 ,bbllx=25pt,bblly=418pt,%
         bburx=272pt,bbury=656pt,width=8cm,height=5cm,clip=}
\vspace{-0.4cm}
\newline\mbox{}{\hskip -1.5 cm}$M_{\gamma\gamma}$ $(GeV)$
\end{center}
\vspace{-0.2cm}
\begin{center}
Fig. 1. $\gamma\gamma$ invariant mass
for $\psi(2S)\rightarrow p\overline{p}\pi^+\pi^-\gamma\gamma$.
\end{center}

{\indent}550 events are obtained for $\psi(2S)\rightarrow 
p\overline{p}\pi^+\pi^-\pi^0$. $\pi^+\pi^-\pi^0$ invariant mass
is shown in Fig. 2. 
At 0.783 GeV is $\omega$ signal.
The mass resolution and efficiency though Monte Carlo simulation  are 
obtained to be 14.6 MeV and
$3.6\%$.

\begin{center}
\parbox[c]{8.0 truecm}{
\epsfig{file=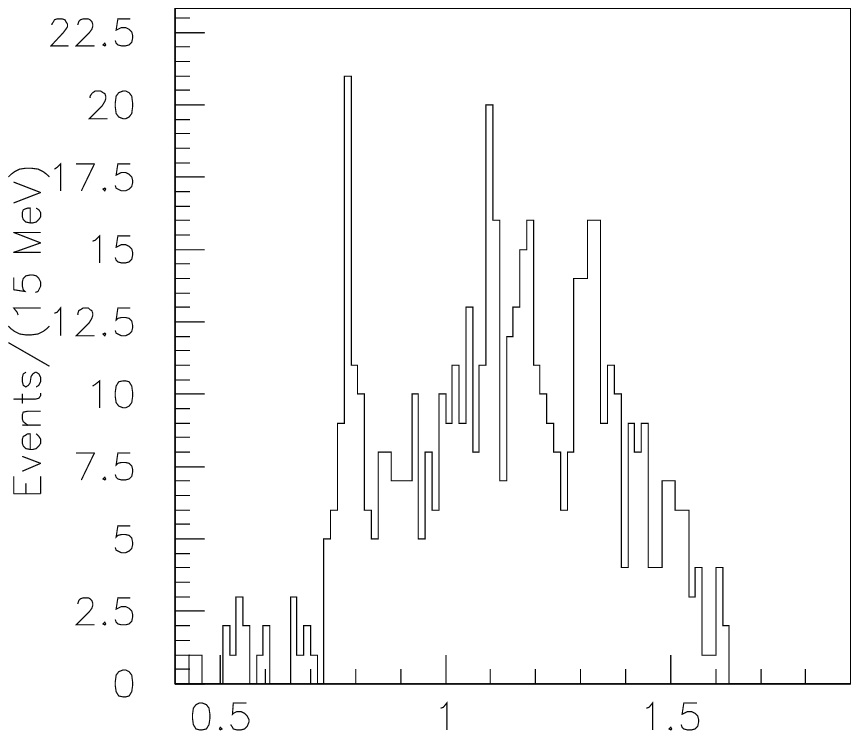 ,bbllx=25pt,bblly=418pt,%
         bburx=272pt,bbury=646pt,width=8cm,height=5cm,clip=}}
\vspace{-0.4cm} 
\newline\mbox{}{\hskip -1.5 cm}$M_{\pi^+\pi^-\pi^0}$ $(GeV)$
\end{center}
\begin{center}
Fig. 2 
$\pi^+\pi^-\pi^0$ invariant 
mass for $\psi(2S)\rightarrow p\overline{p}$ $\pi^+\pi^-\pi^0$.
\end{center}

The $\omega$  mass, width, event number
 can be obtained using
 unbinned Breit-wigner fit. The background is fitted with 
4th order nomial. $M_\omega=783.4\pm4.5\pm0.2$ $MeV$.
$\Gamma_\omega=12.7\pm7.1\pm3.0$ $MeV$.
 $N_{events}=35.0\pm6.0\pm6.7.$

There are not  contriburion to $\omega$ signal from continuum data, which id derived
from $6.42(1\pm4\%)pb^{-1}$ data at 3.65 GeV.

The branching ratio is
$ Br(\psi(2S)\rightarrow p\overline{p}\omega,
\omega\rightarrow\pi^+\pi^-\pi^0)
=\frac{\displaystyle N_{measured}}{\displaystyle \epsilon 
N_{\psi(2S)}}$
$=(6.9\pm1.2\pm1.3)\times 10^{-5}.$
$$Br(\psi(2S)\rightarrow p\overline{p}\omega)=
(7.7\pm1.3\pm1.4)\times 10^{-5}.$$
The BES I data gives$^{[6]}$
$Br(\psi(2S)\rightarrow p\overline{p}\omega)=(8.0\pm3.0\pm 1.0)\times 
10^{-5}.$

{\indent}Now the decay of $\psi(2S)$ to $p\overline{p}\pi^+\pi^-$
is discussed. The charged particle are selected as above.
    To remove the one or  multiple photon events, the missing momentum
is required to be $p_{missing}<0.1$ $GeV$.
 $\psi(2S)\rightarrow
\Lambda\overline{\Lambda}$ is a contamination source.
It is removed by $M_{p\pi^+}>1.15$
$GeV$ and $M_{\overline{p}\pi^-}>1.15$ $GeV$.

Another contamination source is 
$\psi(2S)\rightarrow\pi^+\pi^-J/\psi,$ $J/\psi\rightarrow 
p\overline{p}$. 
It is removed by requiring
$M_{p\overline{p}}<3.0$  $GeV$. 
The $p\overline{p}\pi^+\pi^-$ mass is shown in Fig. 3.
It is $\psi(2S)$ signal. Its resolution is 36 $MeV$.
$\mid M_{p\overline{p}\pi^+\pi^-}-3.686\mid<0.1$ GeV is required.

To remove $\psi(2S)\rightarrow p\overline{p}K^+K^-$
4c fit is performed. 
$\chi^2_{\psi(2S)\rightarrow p\overline{p}\pi^+\pi^-}<30$
and $\chi^2_{\psi(2S)\rightarrow p\overline{p}\pi^+\pi^-}$ 
$<\chi^2_{\psi(2S)\rightarrow p\overline{p}K^+K^-}$ is
required.

\begin{center}
 \psfull
\epsfig{file=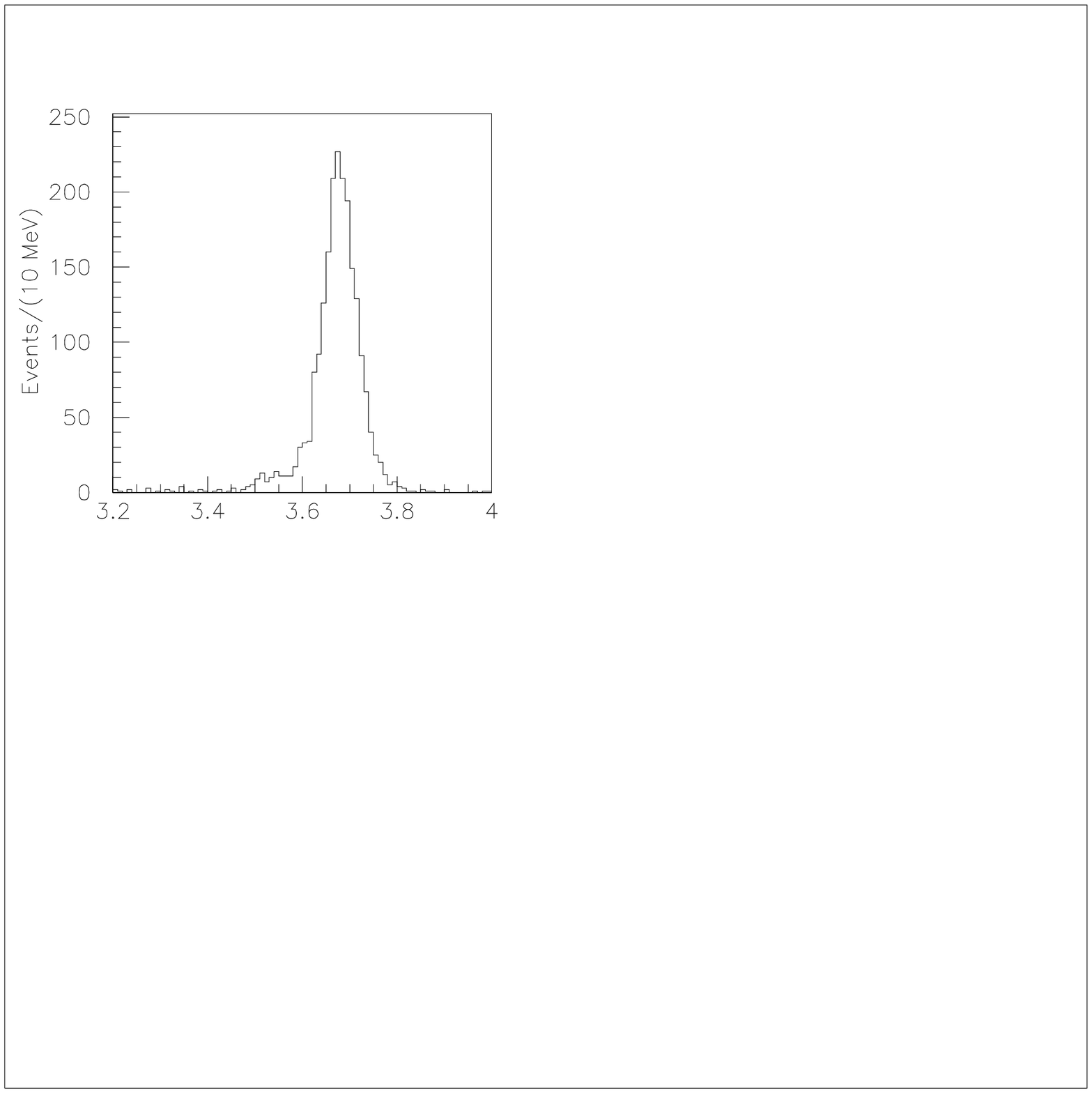 ,bbllx=25pt,bblly=418pt,%
         bburx=272pt,bbury=656pt,width=8cm,height=5cm,clip=}
\vspace{-0.4cm}
\newline\mbox{}{\hskip -1.5 cm}$M_{p\overline{p}\pi^+\pi^-}$ $(GeV)$
\end{center}
\vspace{-0.2cm}
\begin{center}
 Fig. 3. $p\overline{p}\pi^+\pi^-$ invariant mass 
distribution for $\psi(2S)\rightarrow p\overline{p}\pi^+\pi^-$
candidates.
\end{center}

{\hskip 0.4cm}
The $\pi^+\pi^-$ invariant mass  distribution is shown in Fig. 4.
1691 events are there.
Two narrow bumps can be seen at 727.0 and 783.0 MeV and a little
bump at 977.0 GeV, i.e. the position of $f_0(980)$.
From Monte Carlo simulation, the BES-II detector
acceptance efficiencies and  
the mass resolutions at three points are obtained.
The mass resolutions 
are 6.0 MeV at the first two points and 6.8 MeV at 977.0 GeV.
 The efficiencies are
$21.4\%$, $21.8\%$ and $22.4\%$. 
 
Their  masses, widths, event numbers
 and statistic significance (S.S.) can be obtained using
 unbinned Breit-wigner fit. The background is fitted with 3th order nomial.
      Each of the first two bumps is described using free parameters: 
area, width, mass and one fixed parameter: resolution. 
The mass and width of  $f_0(980)$
are fixed at the values of PDG2004 977.0 MeV and 44.0 MeV.
The mass resolution of $f_0(980)$ is fixed.

The table 1
 shows these
 values
\\
\begin{center}
 Table 1. The parameters of bumps by fitting the first two  as unkown 
ones.
\end{center}
\begin{center}
 \begin{tabular}{|c|c|c|c|c|}
 \hline
      &mass (MeV)&  width (MeV)& Event &S.S. $(\sigma)$\\
\hline
bump1 &$726.9 \pm 4.1\pm2.1$&$20.7\pm12.5\pm8.1$ 
&$69.0\pm15.8\pm12.8$&4.9\\
\hline
bump2 &$783.0\pm 2.2\pm 0.4$ &$3.2\pm5.6\pm1.2$ & $42.5 \pm 
10.6\pm 10.7$&5.0\\
\hline
$f_0(980)$&977.0&44.0&$8.8\pm16.4^{+5.6}_{-8.8}$&0.5\\
\hline
\end{tabular}
\end{center}

 statistic significance  is defined as
 $significance=\sqrt{2\times(lnL_{max}-lnL_0)}$
Where $L_{max}$ is likelihood function value
for 1. the first two signals yield with resolutions fixed
and areas, masses and widths to float,
2. the $f_0(980)$ signal yield with resolution, mass, width fixed
and  area to float
and 3. the background shape
parameters to float.
$L0$ is the likelihood function value
for the signal hypothesis under consideration
corresponding to a zero yield and
the other two signals yield as in $L_{max}$.

The mass and width of bump2  are consistent with those 
of $\omega$. 
Bump1 is unknown. 
The known particles listed in 
PDG2004$^{[4]}$ have no consistent masses and widths with those of
bump1. 

The branching ratios are
$$Br(\psi(2S)\rightarrow p\overline{p}bump1,
bump1\rightarrow\pi^+\pi^-)=$$
$$\frac{\displaystyle N_{measured}}{\epsilon N_{\psi(2S)}}
=(2.3\pm 0.5\pm 0.4)\times 10^{-5},$$
$$Br(\psi(2S)\rightarrow p\overline{p}bump2,
bump2\rightarrow\pi^+\pi^-)$$
$$=(1.4\pm 0.4\pm 0.4)\times 10^{-5},$$
$$Br(\psi(2S)\rightarrow p\overline{p}f_0(980),
f_0(980)\rightarrow\pi^+\pi^-)$$
$$=(2.8\pm 5.2^{+1.8}_{-2.8})\times 10^{-6},$$

From $Br(\psi(2S)\rightarrow p\overline{p}\omega,\omega\rightarrow
\pi^+\pi^-\pi^0)$,
it is derived
$Br(\psi(2S)\rightarrow p\overline{p}\omega,\omega\rightarrow
\pi^+\pi^-)=(1.3\pm0.2\pm0.2)\times 10^{-6},$
which is not consistent with branching ratio of bump2 production.

$$Br(\psi(2S)\rightarrow p\overline{p}\pi^+\pi^-)=
\frac{\displaystyle (1691-123-27)\pm40.0\pm175.6}{\displaystyle
0.22\times 14\times 10^6}=(5.0\pm0.1\pm0.6)\times 10^{-4},$$
where $\Lambda\overline{\Lambda}$ is excluded,
 123 is  contriburion of continuum data, which id derived
from $6.42(1\pm4\%)pb^{-1}$ data at 3.65 GeV.
27 is from sidebands. 40.0 is statistical error.
175.6 is systematic error.
The efficiency is $22.0\%$.
The sideband events are obtained using the cut 
$0.1<\mid M_{p\overline{p}\pi^+\pi^-}-3.686\mid<0.2$ GeV 
and the same other cuts for Fig. 4.

In addition to $\Lambda\overline{\Lambda}$,
two decay processes of $\psi(2S)$ have $p\overline{p}\pi^+\pi^-$
final state. One is $\psi(2S)\rightarrow \Delta^{++}\Delta^{--}$
with the branching ratio of $(1.28\pm0.35)\times 10^{-4}$ given
by BES$^{[4,8]}$.
Another  is direct decay $\psi(2S)\rightarrow p\overline{p}
\pi^+\pi^-$. The branching ratio given by PDG2004$^{[4,9]}$ is 
$Br(\psi(2S)\rightarrow p\overline{p}\pi^+\pi^-)=(8.0\pm2.0)\times 
10^{-4}$, which includes 
the contribution from the $\Lambda\overline{\Lambda},$ 
$\Delta^{++}\Delta^{--}$ and direct decay. 

Then what is  
branching ratio of direct decay
$\psi(2S)\rightarrow p\overline{p}\pi^+\pi^-$
from this measurement?
Using $Br(\psi(2S)\rightarrow\Delta^{++}\Delta^{--}),$
Monte Carlo simulation shows that 368 $\Delta^{++}\Delta^{--}$
events contribute Fig. 4, i.e. number of direct decay
$p\overline{p}\pi^+\pi^-$ is 1691-123-27-368-65.9-42.5=1064.6,
two bumps are excluded (because number of $f_0(980)$ is small
and has large statistics, it is ignored). The corresponding 
branching ratio is $3.5\times 10^{-4}$. In Fig. 4
the solid line
is from Monte Carlo
evevts of $\psi(2S)\rightarrow\Delta^{++}\Delta^{--}$
with branching ratio of $1.28\times 10^{-4}$
plus direct decay $\psi(2S)\rightarrow p\overline{p}\pi^+\pi^-$
with branching ratio of $3.5\times 10^{-4}$.

\begin{center}
 \psfull
\epsfig{file=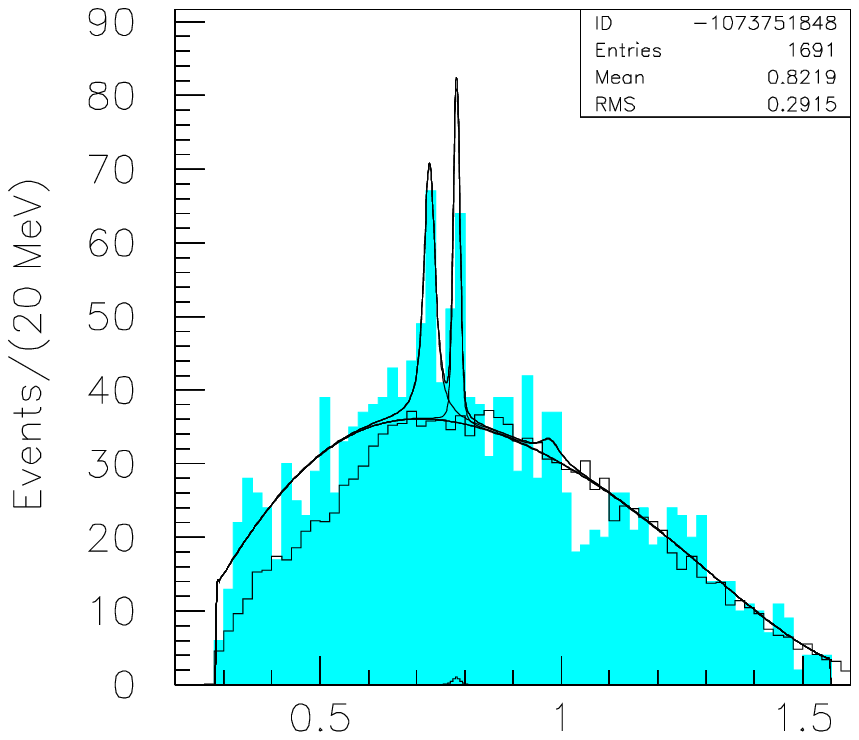 ,bbllx=25pt,bblly=418pt,%
         bburx=272pt,bbury=656pt,width=8cm,height=5cm,clip=}
\vspace{-0.4cm}
\newline\mbox{}{\hskip -1.5 cm}$M_{\pi^+\pi^-}$ $(GeV)$
\end{center}
\vspace{-0.2cm}
{\hskip 0.6cm} Fig. 4. $\pi^+\pi^-$ invariant mass distribution 
 for $\psi(2S)\rightarrow p\overline{p}\pi^+\pi^-.$
 The shaded area is data.
 The smooth line is from BW fit. The solid line is from 
 Monte Carlo simulation
 $\psi(2S)\rightarrow p\overline{p}\pi^+\pi^-$
 and $\psi(2S)\rightarrow \Delta^{++}\Delta^{--}.$
 The small peak at 0.782 GeV near bottom line is from
 Monte Carlo simulation with $Br(\psi(2S)\rightarrow p\overline{p}\omega,
\omega\rightarrow \pi^+\pi^-)=1.3\times 10^{-6}$.

{\vskip 0.4cm}
The sample of $14\times 10^6$ Monte Carlo inclusive $\psi(2S)$ decay
from lund-charm generator$^{[10]}$ is used to estimate the background.
The sample is throught the same cuts.  
 4686 events are selected. Among 
the events,  the contamination is 
shown in table 2.
Monte Carlo simultion for each channel 
shows
these channels maybe have no contributions to $p\overline{p}\pi^+\pi^-$
invaraint mass distribution around  0.727 and 0.783 $GeV$.

\begin{center}
 Table 2. Contamination from $\psi(2S)$ decay to anything
{\vskip 0.2cm}
\begin{tabular}{|c|c|c|}
\hline
source & number& bran. ratio\\
\hline
$\eta^\prime p\overline{p},\eta^\prime \rightarrow
\gamma\rho$&22&\\
\hline
$\eta p\overline{p},\eta\rightarrow 
\gamma\pi^+\pi^-$&13&\\
\hline
$\chi_{c1}\gamma,\chi_{c1}\rightarrow
p\overline{p}\rho$&1&$<4.5\times {10^{-5}}^*$\\
\hline
$\chi_{c2}\gamma,\chi_{c2}\rightarrow
p\overline{p}\rho$&5&$<9.5\times {10^{-5}}^*$\\
\hline
$\chi_{c2}\gamma,\chi_{c2}\rightarrow
a1(1260)^0\pi^0$&1&\\
\hline
$\chi_{c2}\gamma,\chi_{c2}\rightarrow
p\overline{p}\pi^+\pi^-$&1&$9.5\times10^{-5}$\\
\hline
$\pi^0\pi^+\overline{p}\Delta^{--}$&1&\\
\hline
$ p\overline{p}\pi^0$&1&$1.4\times10^{-4}$\\
\hline
$ p\overline{p}\gamma$&2&\\
\hline
$ \pi+\pi^+J/\psi,
J/\psi\rightarrow e^+e^-$&1&$<1.8\times 10^{-2}$\\
\hline
sum&48&\\
\hline
\end{tabular}
\end{center}
here, branching ratios come from PDG2004.
*$Br(\chi_{c1}\gamma,$ $\chi_{c1}\rightarrow
p\overline{p}\rho)<Br(\chi_{c1}\gamma,\chi_{c1}\rightarrow
p\overline{p}\pi^+\pi^-)=4.5\times 10^{-5}$.
$Br(\chi_{c2}\gamma,\chi_{c2}\rightarrow
p\overline{p}\rho)<Br(\chi_{c2}\gamma,\chi_{c2}\rightarrow
p\overline{p}\pi^+\pi^-)=9.5\times 10^{-5}$.

Systematic errors for eficiency are caused by difference between data and 
MC simulation. Studies have determined these errors to be $8\%$  
for the tracking efficeincy, 
$2\%$ for photon identification,
$^{[10]}$, $5.6\%$ for PID,
$1.9\%$, $11.1\%$, $16.3\%$, $-78.6\%$, $8.4\%$
for
 kinematic fit of 
$p\overline{p}\pi^+\pi^-$,
$p\overline{p}bump1,$ $p\overline{p}bump2$,
$p\overline{p}f_0(980)$
 and $p\overline{p}\omega$,
$0\%$, $0.01\%$, $4.1\%$, $62.5\%$ and
 $1.2\%$ for
BW fit of 
$p\overline{p}\pi^+\pi^-$,
$p\overline{p}bump1,$ $p\overline{p}bump2$,
$p\overline{p}f_0(980)$
using 4th order nomial to describe the background
 and $p\overline{p}\omega$ using 5th order nomial
to decribe the background, 
$1.1\%$, $1.7\%$, $8.5\%$, $-12.9\%$ and
 $5.9\%$ for
$\mid cos\theta\mid<0.8$ for
 $p\overline{p}\pi^+\pi^-$,
 $p\overline{p}bump1,$ $p\overline{p}bump2$,
 $p\overline{p}f_0(980)$
 and $p\overline{p}\omega$,
$0.5\%$, $0.5\%$ and $3.1\%$ for effciencies of bump1, bump2 and 
$f_0$, 
$0.6\%$,
$9.7\%$, $10.5\%$ and $-100.0\%$ for
$p_{missing}$ for
$p\overline{p}\pi^+\pi^-$,
 $p\overline{p}bump1,$ $p\overline{p}bump2$
and $p\overline{p}f_0(980)$,
$5.0\%$ for number of $\psi(2S)$ events.
Total systematic errors are 
$11.2\%$, $18.5\%$, $25.2\%$, $^{+63.4}_{-100}$
 and $15.4\%$ for $p\overline{p}\pi^+\pi^-$,
bump1, bump2, $f_0(980)$ and  $\omega$ respectively.


{\indent}
The parameters of the two bumps in Fig. 4 are obtained by fitting them
as completely unkown partilces. But theoretically $\psi(2S)$ can
decay to $p\overline{p}\rho^0$ $(J^{PC}=1^{--})$ and $p\overline{p}\omega$ 
$(1^{--})$. This gives rise to the questions:
1.the $\rho^0$ and $\omega$ plus their interference can fit?
2. the $\omega$ plus one unknown bump1 can fit?
3.  the $\rho^0$ and
$\omega$ and their interference plus one unknown bump1 can fit
the two bumps? 

Fig. 5 showes Breit-Winger fits for these three combinations, in which the
the masses and widthes of the $\rho^0$, $\omega$
and $f_0(980)$ are fixed at PDG2004.
The fitting parameters are given in table 3.

{\small
 \begin{center}
 Table 3. Breit-Winger fitting parameters 
{\vskip 0.2cm}
\begin{tabular}{|c|c|c|c|c|c|c|}
\hline
combination  & bump & mass (MeV) & width (MeV) &event& S.S.& Br.$(10^{-5})$\\
\hline
1&$\rho_0$&776.0&149.0&$156.6\pm31.9\pm36.8$&5.1&$5.4\pm1.1\pm1.3$\\
\cline{2-7}
 &$\omega$&782.5&8.5  &$13.0\pm7.7\pm5.1$&1.8&$0.43\pm0.25\pm0.17$\\
\cline{2-7} 
&$f_0(980)$&977.0&44.0&$20.8\pm16.0\pm9.4$&1.3&$0.66\pm0.52\pm0.30$\\
\hline
2&$\omega$& & &$49.9\pm12.2\pm10.1$&5.0 &$1.6\pm0.4\pm0.3$\\
\cline{2-7}
 &$f_0(980)$& & &$9.5\pm16.4^{+6.0}_{-9.5}$&0.6
&$0.30\pm0.52^{+0.19}_{-0.30}$\\
\cline{2-7}
 &bump1&$726.7\pm4.0\pm2.5$&$19.4\pm10.8\pm7.0$&
$66.5\pm15.6\pm14.9$&4.9&$2.2\pm0.5\pm0.5$\\
\hline
3&$\rho^0$& & & $18.5\pm13.2\pm11.2$&4.5&$6.4\pm4.6\pm3.9$\\
\cline{2-7}
&$\omega$& & &$35.8\pm15.8\pm6.2$&3.8&$1.2\pm0.5\pm0.2$\\   
\cline{2-7}
&$f_0(980)$&&&$20.1\pm15.6\pm9.7$&1.0&$0.64\pm0.50\pm0.31$\\
\cline{2-7}
&bump1&$726.8\pm4.2\pm2.3$&$18.4\pm12.5\pm7.2$&
$63.7\pm14.4\pm11.4$&4.0 &$2.1\pm0.5\pm0.4$\\   
\hline
\end{tabular}
\end{center}

}

{\hskip 1.0cm}
\parbox[c]{4.0truecm}{
\epsfig{file=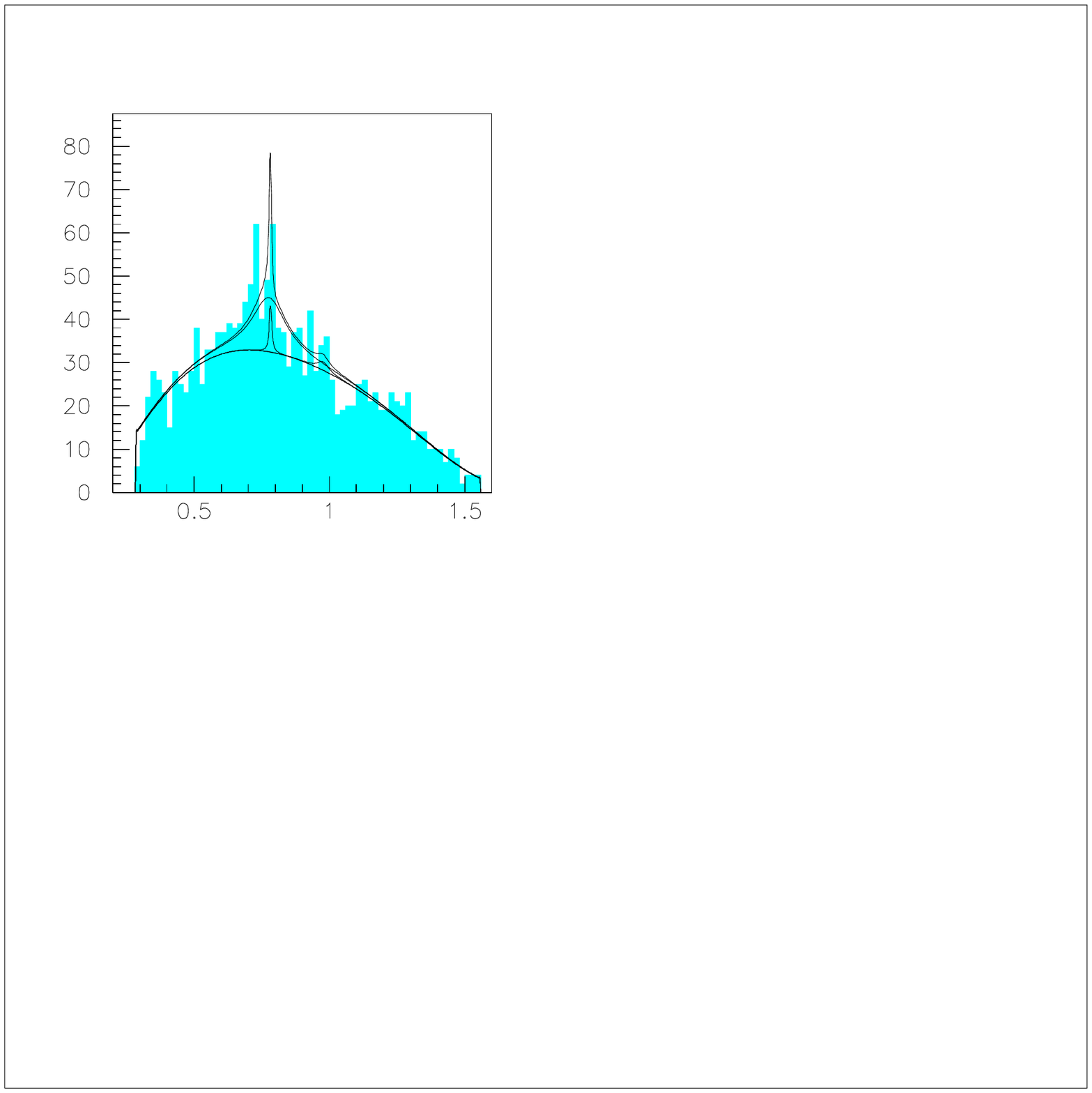 ,bbllx=25pt,bblly=418pt,%
         bburx=272pt,bbury=646pt,width=4cm,height=4cm,clip=}
\vspace{-1.0cm}
\newline\mbox{}{\hskip 0.7cm} (1).$M_{\pi^+\pi^-}$ $(GeV)$
}   
{\hskip 1.0 cm}
\parbox[c]{4cm}{
\epsfig{file=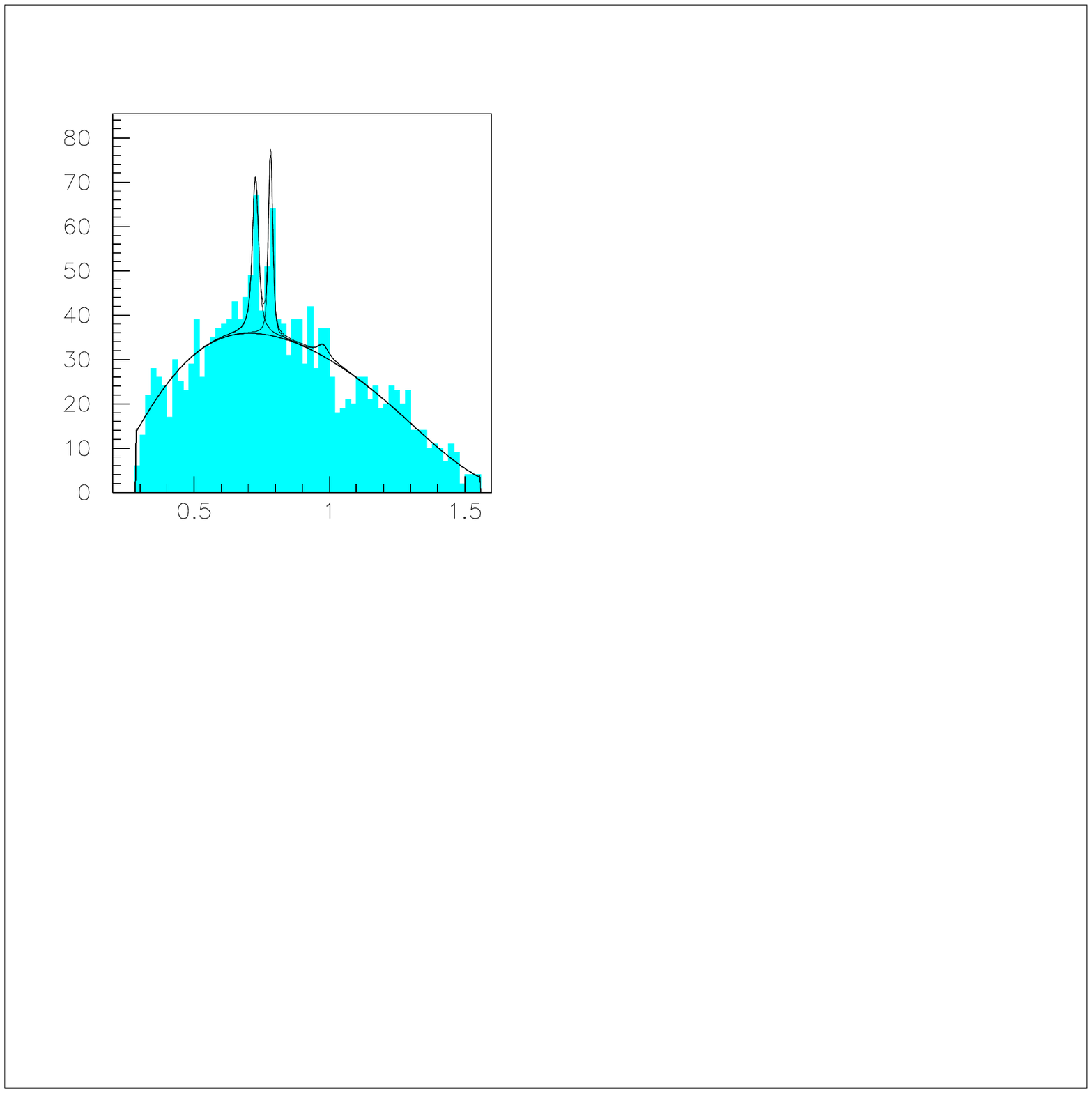 ,bbllx=25pt,bblly=418pt,%
         bburx=272pt,bbury=646pt,width=4cm,height=4cm,clip=}
\vspace{-1.0cm}
\newline\mbox{}{\hskip 0.8cm}(2).$M_{\pi^+\pi^-}$ $(GeV)$
}
{\hskip 1.0 cm}
\parbox[c]{4.0truecm}{
\epsfig{file=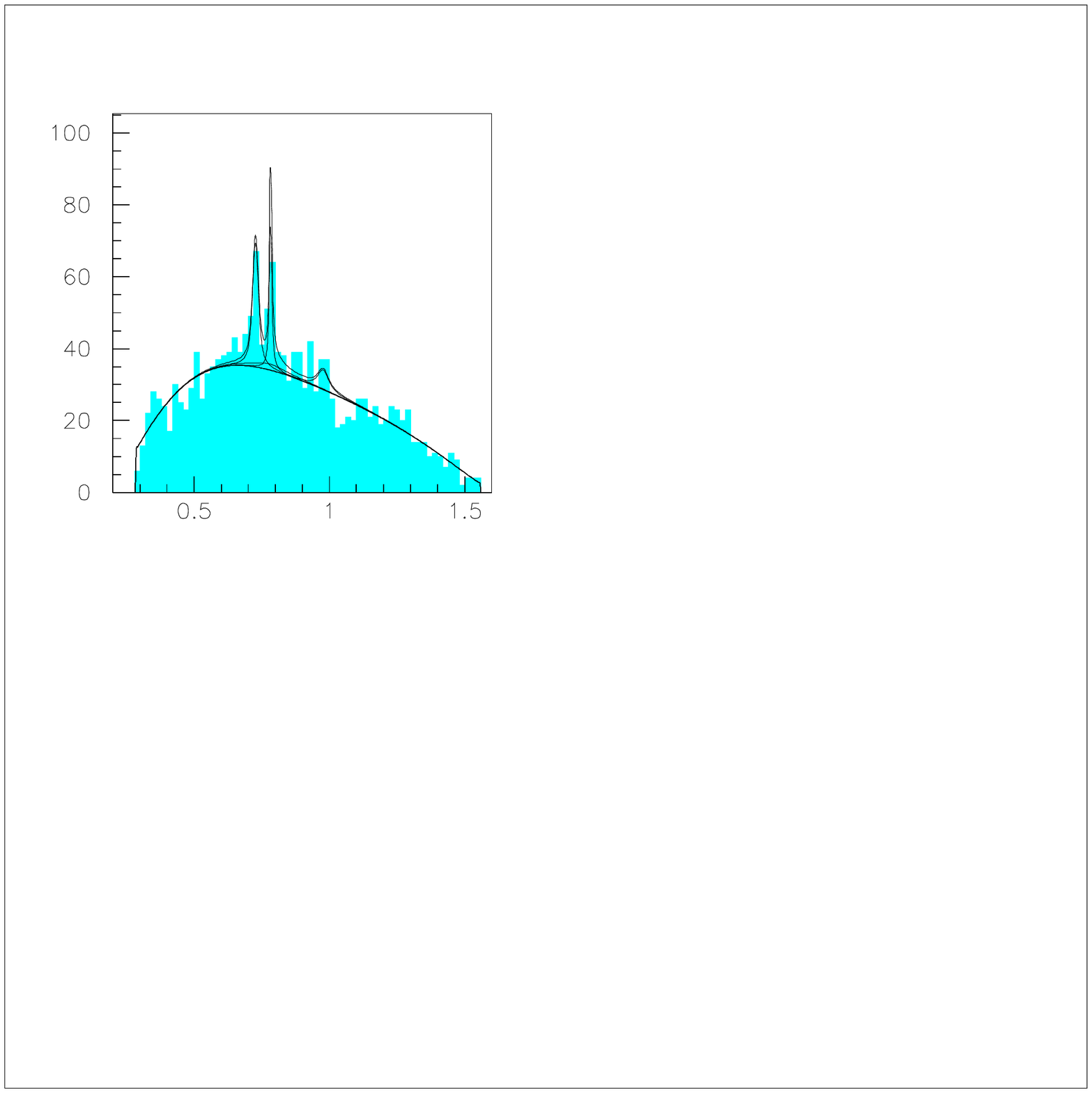 ,bbllx=25pt,bblly=418pt,%
         bburx=272pt,bbury=646pt,width=4cm,height=4cm,clip=}
\vspace{-1.0cm}
\newline\mbox{}{\hskip 0.8cm}(3).$M_{\pi^+\pi^-}$ $(GeV)$
}
{\vskip 0.2cm} 
Fig. 5  Three kind of combinations of  $\rho^0$
 $\omega$  $f_0(980)$ and bump1  to fit the $\pi^+\pi^-$ invariant mass 
distribution from $\psi(2S)\rightarrow p\overline{p}\pi^+\pi^-$.
(1)$\rho^0~+~\omega~+\rho^0\omega ~interference~+~f_0(980)$. 
(2)$unknown~bump1~+~\omega~+~f_0(980)$.
(3)$unknown~bump1~+~\rho^0~+~\omega~+~\rho\omega~interference~+~f_0(980)$.
\\
\\
In this work,
$br(\psi(2S)\rightarrow p\overline{p}\pi^+\pi^-)=(5.0\pm0.1\pm0.6)\times 10^{-4}$ and
$br(\psi(2S)\rightarrow p\overline{p}\omega)=(7.7\pm1.3\pm1.4)\times 10^{-5}$
are measured by analysis of BES-II 14 million $\psi(2S)$ events sample. This leads to
the observation of two  bumps   in $\pi^+\pi^-$ mass
spectrum in  the decay process
$\psi(2S)\rightarrow p\overline{p}\pi^+\pi^-$.
The bumps are not observed in $J/\psi\rightarrow 
p\overline{p}\pi^+\pi^-$$^{[4]}$. 
The mass and width of the first bump can not match those of any 
particles in PDG2004. Another bump 
has a consistent mass and width with those of $\omega$.
But if bump2 is real $\omega$, 
there is no consistency between the
branching ratio 
for $\psi(2S)\rightarrow p\overline{p}bump2,
bump2\rightarrow\pi^+\pi^-$  
and the branching ratio
for $\psi(2S)\rightarrow p\overline{p}\omega,
\omega\rightarrow\pi^+\pi^-\pi^0.$

Duo to small statistics, the quantum number $J^{PC}$ 
is not set for bump1 and bump2. The bump1 has a statistical significance
$4.0< S.S. <4.9$ $\sigma$. Whether the two bumps come from statistical
fluctuation should be confirmed by larger statistics of $\psi(2S)$ events.

The BES collaboration thanks the staff of BEPC for their hard
efforts. This work is supported in part by the National Natural
Science Foundation of China under contracts Nos. 10491300,
10225524, 10225525, 10425523, the Chinese Academy of Sciences under
contract No. KJ 95T-03, the 100 Talents Program of CAS under
Contract Nos. U-11, U-24, U-25, and the Knowledge Innovation
Project of CAS under Contract Nos. U-602, U-34 (IHEP), the
National Natural Science Foundation of China under Contract No.
10225522 (Tsinghua University), and the Department of Energy under
Contract No.DE-FG02-04ER41291 (U Hawaii).


\begin{thebibliography}{99}
\bibitem{1}A. M. Boyarski {\it et al}, Phys. Rev. Lett. 34(175) 1357. 
\bibitem{2}V. Luth {\it et al}., Phys. Rev., Let. 25(1975)1124.
\bibitem{3} M. W. Eaton {\it et al},Phys. Rev. D29(1984)804.
\bibitem{4}Particle Data Group, Phys. Lett. 592 (2004) 1.
\bibitem{5}B. N. Ratcliff {\it et al}, Phys. Lett. B38(1972)345.
\bibitem{6}J .Z. Bai {\it et al.}, Phys. Rev. D63(2001)032002.
\bibitem{7}J .Z. Bai {\it et al.}, Nucl. Inst. Meth., A344 (1994) 319.
\bibitem{8}J .Z. Bai {\it et al.}, Phys. Rev. D67(2003)052002 
\bibitem{9} W. Tanenbaum {\it et al.}, Phys. Rev. D17(1978) 1731.
\bibitem{10}C. J. Chen {\it et al.}, Phys. Rev. D62(2000)034003
\end{thebibliography}
\end{document}